\documentclass{article}
\usepackage{epsfig}
\usepackage{graphicx}
\begin{document}
\begin{center}
{\bf \large W physics
at the ILC with polarized beams as a probe of
the Littlest Higgs Model}
\vskip 1cm
{\bf B. Ananthanarayan\\
Monalisa Patra}\\
\medskip
Centre for High Energy Physics\\
Indian Institute of Science\\
Bangalore 560 012, India\\
\bigskip
{\bf P. Poulose}\\
\medskip
Department of Physics\\
Indian Institute of Technology Guwahati\\
Guwahati 781 039, India\\
\bigskip

\end{center}
\begin{abstract}
We study the possibility of using W pair production and leptonic
decay of one of the W's at the ILC with polarized beams as a probe
of the Littlest Higgs Model.  We consider cross-sections,
polarization fractions of the W's, leptonic decay energy and
angular distributions, and left-right polarization asymmetry 
as probes of the model.  
With parameter values allowed by present
experimental constraints detectable effects on these observables
at typical ILC energies of 500 GeV and 800 GeV will
be present. Beam polarization is further found to enhance the sensitivity. 
\end{abstract}

\section{Introduction}
One of the important processes that will be studied at high
precision at ILC with and without beam 
polarization is W-pair production.  
Phenomelonogical studies of this process within the 
Standard Model (SM) have been carried out in great detail 
~\cite{Hagiwara,Gounaris}.  
Since properties of the weak gauge bosons are closely linked to electroweak 
symmetry breaking (EWSB) and the structure of the gauge sector 
in general, detailed study of W physics
will throw light on what lies beyond the SM. 
The study of mechanisms of EWSB is one of the main concerns of 
particle physics today. The standard Higgs mechanism is less than 
satisfactory, and faces difficulties such as the hierarchy problem. Looking 
beyond the SM, the newly proposed Little Higgs scenarios 
~\cite{lhintro, lsth}
provide a dynamical
way to generate the EWSB, in contrast to the 
{\it ad hoc} introduction of the 
elementary scalar sector. Apart from this 
aesthetically appealing feature, Little Higgs models provides rich 
phenomenology with predictions that could be vidicated or ruled out at
future colliders such as the LHC and the ILC.

One major feature of such models is the presence of additional gauge bosons
in the physical spectrum.  These influence processes like
W-pair production in $e^+e^-$ collisions, firstly directly through their
exchange in the process, and secondly through the
change of standard couplings
through mixing with other gauge bosons.
Although these additional gauge bosons are typically too heavy
to be produced at reference ILC energies of 500 and 800 GeV which we
use in the present study, their effects manifest themselves
as stated above.
Recently it was pointed out, in
a preliminary study, that for one such
model known as the Littlest Higgs Model (LHM), the fraction of
longitudinally and transversely polarized of
one of the W's could be significantly
different from the corresponding fraction in the SM~\cite{Poulose}.  
In this work, we consider a refined treatment of
the LHM to be described below, and extend the prior work to
polarization fractions and total cross-sections of the W's, 
energy and angular distributions of decay leptons, as well as to
observables like the forward-backward
and left-right asymmetries, which are more sensitive
to the effects of the LHM compared to the cross-sections. 

The ILC is expected to have large beam polarizations which
will significantly enhance the sensitivity to new physics, for
a review, see ref.~\cite{hep-ph/0507011}.  We consider
different beam polarizations with the aim of improving the sensitivity of
the observables considered here.

This article is organized as follows:
in Section 2 we, very briefly, introduce the LHM, and
describe its particle spectrum and couplings relevant to $e^+e^-\rightarrow
W^+W^-$.
In Section 3 we present our analysis of the 
total cross section and W-polarization fractions in the LHM
and compare with the SM case. 
In Section 4 
we take up the task of probing the model by considering decay of
one of the $W's$ to a lepton pair.
We consider the energy and angular distributions for the cases of SM and
LHM. We also discuss the left-right as well
as forward-backward asymmetry in this
section. Finally we summarize our study and present our conclusions in
Section 5.  Note that we have included beam polarization effects
in this study in each of the relevant sections.

\section{The Littlest Higgs Model and $W$ Pair Production at ILC
}

In Little Higgs models \cite{lhintro} a non-linear realization of some global 
symmetry {$G$ broken down to $H$} is considered. The Nambu-Goldstone Bosons 
(NGB) of the symmetry
breaking are candidate Higgs fields.  In a specific model, called LHM 
\cite{lsth} $G\equiv SU(5)$ is broken down to
$H\equiv SO(5)$ via a 
vacuum expectation value (vev) of order $f$.
Interactions of NGB's are described by a non-linear sigma model, which
is an effective theory valid below the cut off
{$\Lambda \sim 4\pi\:f$}.
In the version of the LHM \cite{lsth, lsthG,hep-ph/0303236} we will consider 
in this report, 
$SU_1(2)\times SU_2(2)\times U_Y(1)~\subset~ SU(5)$ is gauged, 
which is broken down to the SM gauge group 
{$SU(2)_L\times U(1)_Y$}. 
Under this, the 14 NGB's transform as a real singlet, a real 
triplet, a complex doublet and a complex triplet.
The real fields become the longitudinal degrees of the heavy gauge bosons, 
while the SM gauge bosons, $\vec{W}_L^\mu$ and ${B}_L^\mu$  remain massless at
this stage. 
The doublet NGB field has the correct quantum numbers to
be identified as the standard Higgs doublet. At tree level, they have only
derivative couplings, but quantum corrections at one-loop level generate a
Coleman-Weinberg potential with quadratic and quartic terms, consequently
breaking electroweak symmetry.
Gauge symmetry is constructed such that,
in the absence of any one (original) gauge interaction the
Higgs is massless to all orders. This also ensures that quadratically
divergent contributions to the mass-square term at one-loop level are
cancelled between the gauge bosons from the two sectors.
Logarithmically divergent terms contribute to the potential.
In order to avoid a quadratic divergence due to a
top-quark loop, a pair of (weak-singlet) Weyl quarks
{$U_L,\; U_R$} is introduced, which mix with the ordinary left-
and right- quarks to give mass eigenstates.
Here again, it is so arranged such that the quadratic divergence coming from
the standard top-quark is cancelled by its heavy counterpart, and the
logarithmically divergent part is added to the Coleman-Weinberg potential.
The model achieves EWSB, at the same time, protecting the Higgs mass from 
accquiring quadratically divergent corrections at one loop.  It
is therefore to be expected that these models will have a rich phenomenology
with distinct signatures that can be probed at upcoming collider
experiments. 
An incomplete list of phenomenological studies of different variations of 
the Little Higgs Model 
scenarios 
is ref.~\cite{lhpheno, hep-ph/0211124, hep-ph/0303236, Dobado}.

\begin{figure}[htb]
\begin{center}
\includegraphics[width=10 cm,height=25mm]{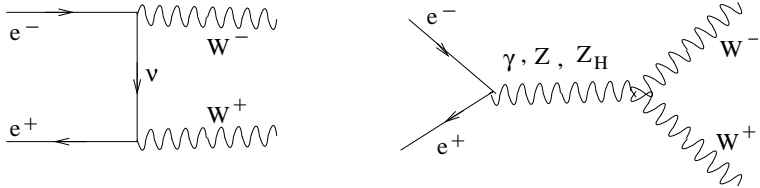}
\caption{Feynman diagrams contributing to the process $e^-e^+\rightarrow
W^+W^-$ in the LHM.}
\label{fig:FD}
\end{center}
\end{figure}

Our interest here is the effective theory below the cut-off $\Lambda$. 
For the process $e^+e^-\rightarrow W^+W^-$, 
we have an $s-$channel process with the
exchange of the heavy neutral gauge boson, $Z_H$, in addition to the standard 
channels as shown in Fig.~\ref{fig:FD}.

Apart from the contribution due to the additional $s$-channel process, the 
LHM also changes the SM couplings. The
relevant couplings in terms of the parameters of the LHM is given
below in terms of 
the global symmetry breaking scale $f$, the parameter $\cos\theta$, which 
represents mixing between the two gauge sectors of the LHM
and the vev. 

The three-point gauge couplings involving $WW$ are given by:
\[
V^\mu (k_1)W^\nu (k_2)W^\rho (k_3)=ig_{VWW}~\left[
g^{\mu\nu}(k_1-k_2)^\rho+
g^{\nu\rho}(k_2-k_3)^\mu+
g^{\rho\mu}(k_3-k_1)^\nu \right],
\]
where all the momenta are considered outflowing, and $V\equiv \gamma,~Z,~Z_H$.
Individual $g_{VWW}$ are given by:

\begin{eqnarray}
&&g_{\gamma WW}=-e\\
&&g_{ZWW}=-e~\frac{\cos\theta_W}{\sin\theta_W}\\
&&g_{Z_{H}WW}=\frac{ev^2}{8f^2\sin \theta_W}~\sin4\theta
\end{eqnarray}

Fermion couplings are given by:

\begin{eqnarray}
&&g_{e\nu W}= i\frac{g}{2\sqrt{2}}
\left[1-\frac{v^{2}}{2f^{2}}\cos^{2}\theta\cos2\theta\right]~
\gamma^{\mu}~(1-\gamma^5)\\
&&g_{eeV}=i\gamma^{\mu}(c^{v}_V-c^{a}_V\gamma^{5}),
\end{eqnarray}
where
\begin{eqnarray}
&&c^{v}_{\gamma}=-e;~~~~~~~~~~~~~~~~~ c^{a}_{\gamma}=0\\
&&c^{v}_{Z}=-\frac{e}{\sin2\theta_W}\left[\left(-\frac{1}{2}+2\sin^{2}\theta_W\right)-
\frac{v^{2}}{f^{2}}\frac{\sin4\theta\cot\theta}{2}\right]\\
&&c^{a}_{Z}=
-\frac{e}{\sin2\theta_W}\left[-\frac{1}{2}-\frac{v^{2}}{f^{2}}
\frac{\sin4\theta\cot\theta}{2}\right]\\
&&c^{v}_{Z_{H}}=c^{a}_{Z_{H}}=-\left(\frac{e\cot\theta}{4\sin\theta_W}\right)
\end{eqnarray}

Following ref.~\cite{hep-ph/0211124} we consider the measured 
values of the Fermi coupling constant, $G_F$, the $Z$-boson mass, $M_Z$, and 
the fine structure constant, $\alpha_{em}(M_Z^2)$ as the Standard Model 
input parameters. The weak mixing angle is obtained from the relation:
\[ \sin^2\theta_0~\cos^2\theta_0=\frac{\pi\alpha_{em}(M_Z^2)}{\sqrt{2}
G_FM_Z^2}\].

The bare weak mixing angle, $\theta_W$ is related to the measured weak coupling angle, $\theta_0$ through the following relation:
\[
\cos \theta_W=\cos\theta_0\left[
1+\frac{\sin^2 \theta_{0}}{\cos^2 \theta_{0}-\sin^2 \theta_{0}}
\left(\frac{v^{2}\cos^{2}\theta\sin^{2}\theta}{2 f^{2}}+
2\frac{|v'|^2}{v^2}\right)\right]
\]
The weak coupling constant expressed in terms of the other parameters becomes
\[
 g=\frac{2 m_{W}}{v}\left[1+\frac{v^{2}}{2f^{2}}\left(\frac{1}{6} +
\frac{(\cos^{2}\theta-\sin^{2}\theta)^{2}}{4}\right)-
2\frac{|v'|^2}{v^2}\right].
\]

The triplet vev, $v'$ is related to the generated quartic coupling, and the 
guage couplings through
\[
\frac{|v'|^2}{v^2}=\frac{v^2}{144f^2}\left(1+\frac{6\lambda-4ag_1^2}
{a(2g'^2+g_1^2)}\right)^2,
\]
which is further constrained to ${|v'|^2}/{v^2}<{v^2}/{16f^2}$ 
\cite{hep-ph/0303236}.
In our numerical analysis we consider the approximate relation,
${|v'|^2}/{v^2}={v^2}/{144f^2}$. Note that our numerical results are
not very sensitive to this choice.  Thus we are left with two free parameters
$f$ and $\theta$. As argued by \cite{Dobado}, precision electroweak 
measurements restrict the parameters to be $f\sim 1$ TeV and 
$0.1<\cos\theta<0.9$. 
In our numerical analysis we consider some representative
values satisfying these restrictions.

\section{Analyses of $e^-e^+\rightarrow W^+W^-$}
In this section we present the results of our numerical analysis to probe the 
LHM through the process $e^-e^+\rightarrow W^+W^-$ at the ILC. 

\subsection{The total cross section}
We compute the total cross section incorporating beam polarization 
using the helicity amplitudes given in ref.~\cite{Hagiwara} with the new 
couplings and with the added contribution due to the exchange of $Z_{H}$.
With beam polarization, in general,
the polarized cross section may be expressed
as~\cite{Pankov:1994hx}:
\begin{eqnarray}
\sigma(e^{+}e^{-}\rightarrow W^{+}W^{-}) &=&\frac{1}{4}\left[(1+P_{e^{-}}).(1-P_{e^{+}})\sigma^{RL}\right. \nonumber \\
    &&+\left.(1-P_{e^{-}}).(1+P_{e^{+}})\sigma^{LR}\right],
\end{eqnarray}
where \(~~
\sigma^{RL}=\sigma(e^{+}_Le^{-}_R\rightarrow W^{+}W^{-}) ~~{\rm and}~~
\sigma^{LR}=\sigma(e^{+}_Re^{-}_L\rightarrow W^{+}W^{-}), 
\) with
$e_{L,R}$ representing the left- and right-polarized electrons 
(and positrons), respectively. The degree of polarization is defined 
as: 
$P_{e}=(N_R-N_L)/(N_R+N_L)$, where $N_{L,R}$ denote the number 
of left-polarized and right-polarized electrons (and positrons), respectively.
More than 80\% of electron beam polarization and large positron beam 
polarization are expected to be achieved at ILC. In our analysis we consider
the ideal possibility of 100\% polarization of the beams.  Our results
are presented in the figures below.

In Fig.~\ref{fig:cs} we present
the total production cross-section for a typical choice of
parameters of the LHM and in the SM of the case of
unpolarized and polarized beams with a specific choice
of beam polarization.
It may be seen that
in the case of unpolarized beams the cross section of LHM 
does not deviate much from that of the SM for energies up to 1 TeV. 

The presence of beam polarization changes this situation significantly. 
The combination, $P_{e^{-}}=+1$ and $P_{e^{+}}=0$ is seen to provide
the largest 
deviation, which is the case we have chosen to display. 
However, for this configuration a reduced number of $W$-pairs is 
produced, as the dominant $t$-channel is cut off. 
Notice also that there is no contribution
due to the $Z_{H}$ exchange in this case, as the $Z_H$ couples only to the 
left-handed electrons.  The effect, therefore is purely due to the deviation
of the standard model couplings.  We will return to some more
properties of this in the next sub-section. 
\begin{figure}[htb]
\includegraphics[width=6 cm,height=6 cm]{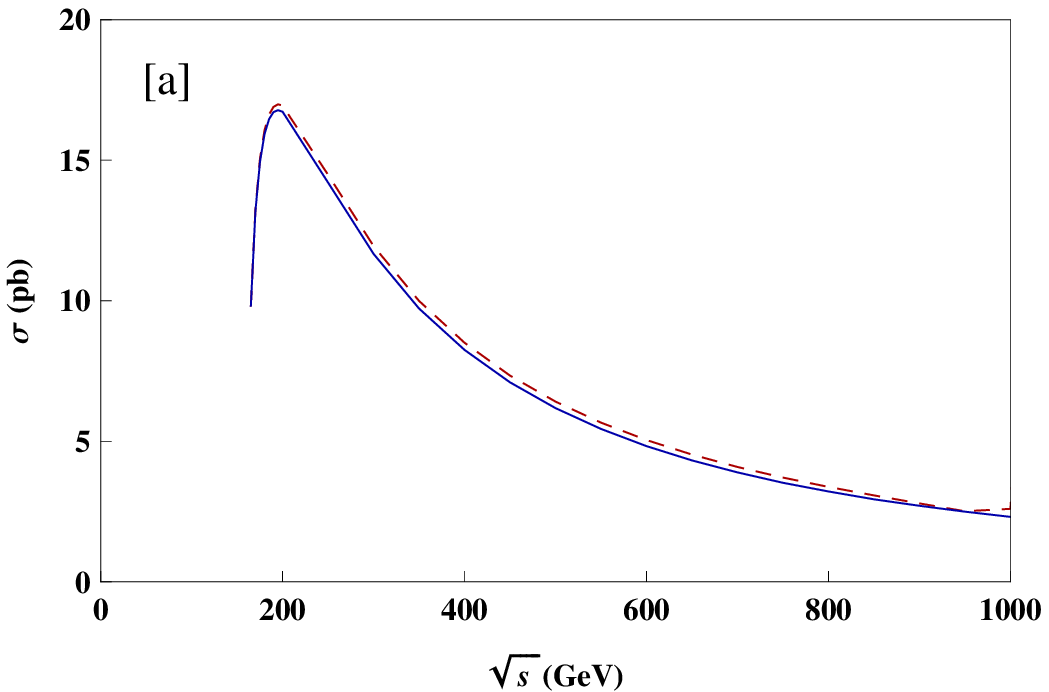}
\hspace{0.2cm}
\includegraphics[width=6 cm,height=6 cm]{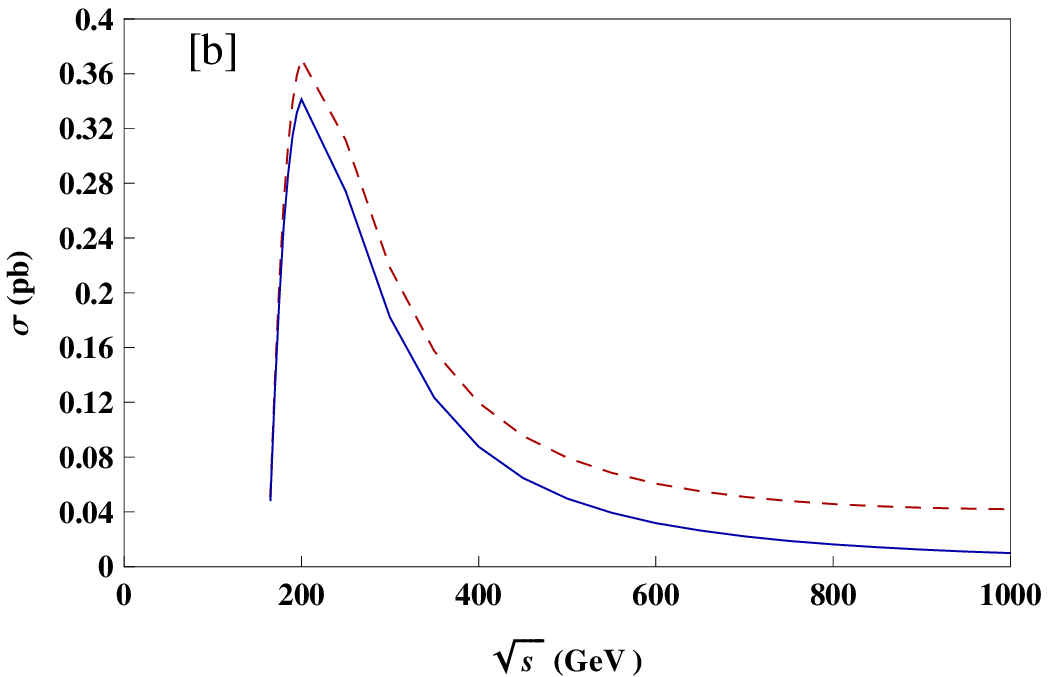}
\caption{Total cross section for $W^{+}$ $W^{-}$ production in 
an $e^{+}e^{-}$ collision for SM [blue-solid] and LHM 
[red-dashed] with [a] unpolarized beams [b] polarized beams with 
$P_{e^{-}}$=1 and $P_{e^{+}}$=0.
The parametres for LHM are $f$=1 TeV and $c$=0.3 }
\label{fig:cs}
\end{figure}
\noindent

\subsection{$W$ Polarization Fractions}
Here we explore the sensitivity of the ILC to the LHM when we
consider polarization fractions of the W bosons.  Such measurements
have been considered in past experiments for precision studies
of the W boson properties at LEP, which 
has measured the fractional cross section of the polarized
$W$'s~\cite{WPfrac}.
At ILC higher precision is expected to be reached.
We define the polarization fractions as
\begin{equation}
f^{0}\equiv \sigma (e^{+}e^{-}\rightarrow W^{-}_{L}W^{+})/\sigma_{unpol}
\end{equation}
\begin{equation}
f^{T}\equiv\sigma (e^{+}e^{-}\rightarrow W^{-}_{T}W^{+})/\sigma_{unpol},
\end{equation}
where $L$ stands for longitudinal polarization, and $T =\pm$ stands for 
transverse polarizations. 

\begin{figure}[htb]
\includegraphics[width=6 cm,height=6 cm]{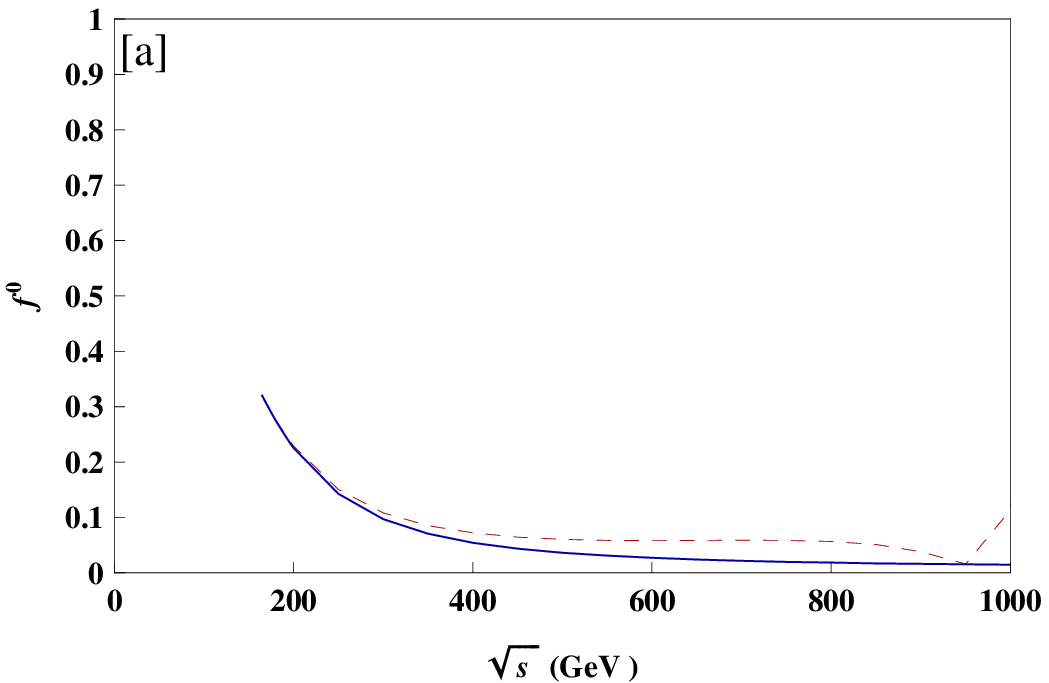}
\hspace{0.2cm}
\includegraphics[width=6 cm,height=6 cm]{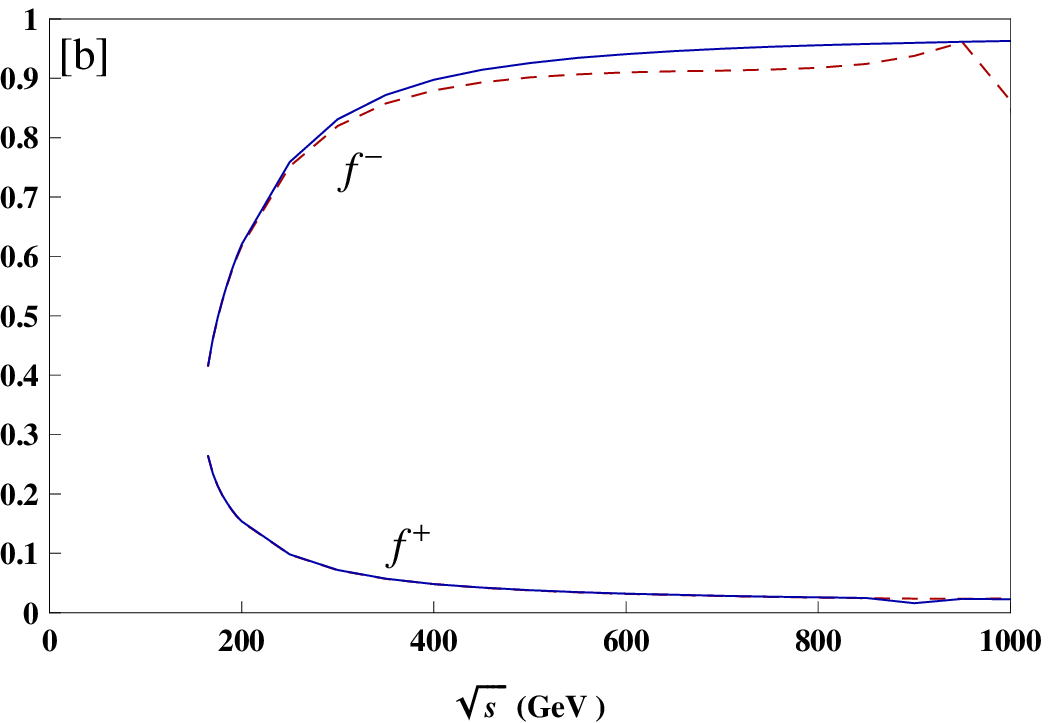}
\caption{Fractional cross section of $W^{-}$,  [a] longitudinal ($f^0$),
and [b] transverse ($f^\pm$),
with unpolarised beams for SM [blue - solid] and LHM [red - dashed].
The parametres in this case are $f$=1 TeV and $c$=0.3}
\label{fig:f}
\end{figure}

The three polarization fractions are studied as a function
of $\sqrt{s}$ which are plotted in Fig.~\ref{fig:f}. 
It is readily observed that
there is a significant deviation in the case of $f^0$ and
$f^-$, while $f^+$ is largely unaffected. 
Since these fractions depend on 
various couplings in a complicated way, 
we do not attempt to explain the
effects in terms of the changes in the couplings. 

In Table ~\ref{tab:sig_f} we present 
firstly the ratio of the cross section
in the LHM to that in the SM for
typical parameter choices, as well as
the ratio of $f^0$ in LHM to that in the 
SM at $\sqrt{s}=500$ GeV and 800 GeV for the two illustrative parameter 
space points. At $\sqrt{s}$=500 GeV for $f$=1 TeV and $c$=0.3 the deviation is 
about 65\% with unpolarized beams. This is improved marginally to about 66\% 
for $P_{e^-}=-1,~~P_{e^+}=0$. While for the slightly larger value of, 
$f=1.5$ TeV, the effect is not as dramatic, we still have significant
deviation of about 25 - 30\% in $f^0$. 
On the other hand, the configuration with purely right handed electrons
and unpolarized positrons has the effect of completely washing out
the effect of the LHM in the polarization fractions.  Therefore,
beam polarization has the dramatic effect of disentangling the
effects of the new physics.  Crucially, there is an interplay between
the pure left-handed coupling nature of $Z_H$ and the corrections
to the couplings of the SM Z-boson coming from the parameters of
the LHM which makes this possible.
Considering the precision at which 
$f^0$ could be measured at ILC, such effects are very interesting.

\begin{table}
\begin{center}
\begin{tabular}{|c|c|l|c|c|c|c|r|} \hline
&&&&\multicolumn{2}{|c|}{}&\multicolumn{2}{|c|}{}\\
 & & & &\multicolumn{2}{|c|}{$\sqrt{s}$ =500GeV}&
\multicolumn{2}{|c|}{$\sqrt{s}$ =800GeV} \\[2mm] \cline{5-8}
&&&&&&&\\
$P_{e^-}$ &$P_{e^+}$& $f$ (TeV)&$c$
&$\sigma_{LHM}/\sigma_{SM}$& $f^0_{LHM}/f^0_{SM}$
&$\sigma_{LHM}/\sigma_{SM}$& $f^0_{LHM}/f^0_{SM}$\\[2mm] \hline\hline
  &  & 1 &0.3   &1.04&1.65&1.05&3.01            \\
0 &0 &1.5&0.3   &1.02&1.27&1.03&2.06                \\
  &  &1.5&0.5   &1.00&1.25&1.00&1.72                \\ \hline
  &  & 1 &0.3   &1.04&1.66&1.05&3.08            \\
-1&0 &1.5&0.3   &1.01&1.28&1.02&2.12                \\
  &  &1.5&0.5   &1.00&1.26&1.00&1.75                \\ \hline
  &  & 1 &0.3   &1.60&1.00&2.81&1.00            \\
+1&0 &1.5&0.3   &1.25&1.00&1.69&1.00                \\
  &  &1.5&0.5   &1.17&1.00&1.46&1.00                \\ \hline
\end{tabular}
\end{center}
\caption{Ratios of the cross section and 
W polarization fractions in the LHM to those in
the SM for different beam polarizations, and for some illustrative
values of $f$ and
$c$.}
\label{tab:sig_f}
\end{table} 
\subsection{Angular Spectrum of the Secondary Lepton}
In order to exploit further the process at hand, it is profitable
to consider the decays of one or both the W's.  
Let us consider
$e^{+}e^{-}\rightarrow W^{+}W^{-}$ with 
$ W^{-}\rightarrow l^{-}\bar{\nu }$ and $W^+$ going into anything.
Energy-angle correlation of the secondary leptons is given by the following
expression~\cite{Kov, BESS}: 
\begin{eqnarray}
\frac{d\sigma}{dx~d\cos\theta_{l}}&=&\frac{3}{2}\frac{\alpha^{2}}{s}~BR(W^{-}\rightarrow e^{-}\bar{\nu})~A(s,x,\theta_{l}) \nonumber \\
                    && \times\left[\arctan\left(\frac{m_{W}}{\Gamma_{W}}\right)+\arctan\left(\frac{sx}{m_{W}\Gamma_{W}}-\frac{s\tau}{m_{W} \Gamma_{W} (1-x)}\right)\right]. \nonumber\\
\end{eqnarray}
Expression for the function $A(s,x,\theta_{l})$ is given in the Appendix 
for arbitrary beam polarization, while in the above
$x={2E_l}/{\sqrt{s}}$, where $E_l$ is the energy of the secondary
lepton in the $e^+e^-$ centre of mass frame with $\sqrt{s}$ the centre of mass 
energy, and $\theta_l$ is its polar angle.
$BR(W^{-}\rightarrow e^{-}\bar{\nu})$ is the leptonic branching ratio of $W$,
$\Gamma_W$ is its width, and $\tau={m_W^2}/{s}$.
In principle, the decay width and branching ratios of $W$ can be different
from those of the SM values. But, in the case when the additional fermions
are heavy, we can assume an SM like decay of the $W$. In our analysis we have
taken this approach.

\begin{figure}[htb]
\includegraphics[width=6 cm,height=5.2 cm]{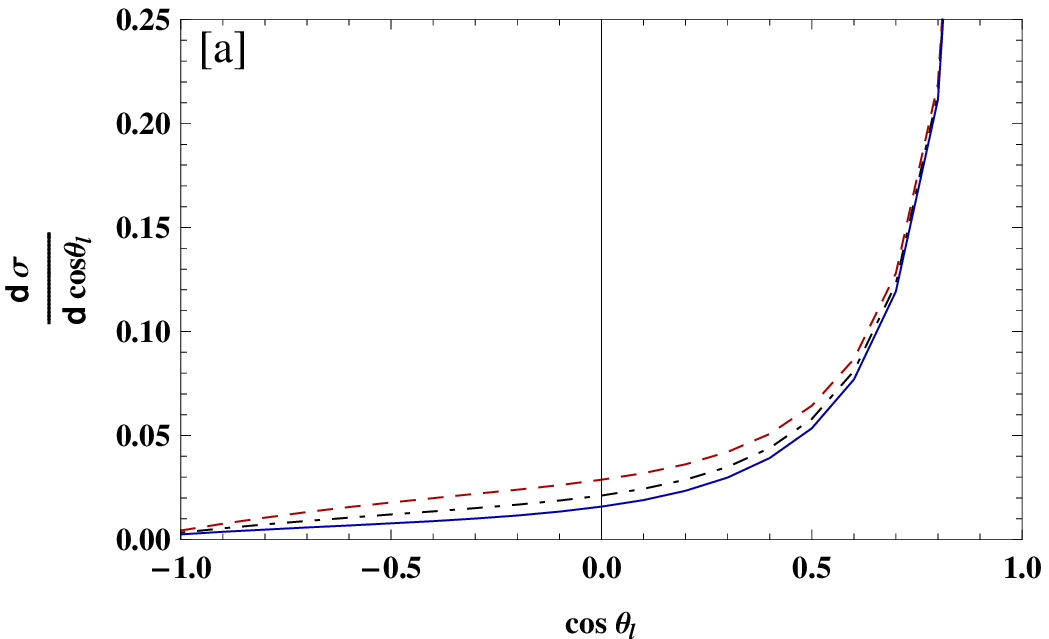}
\hspace{0.2cm}
\includegraphics[width=6 cm,height=5.2 cm]{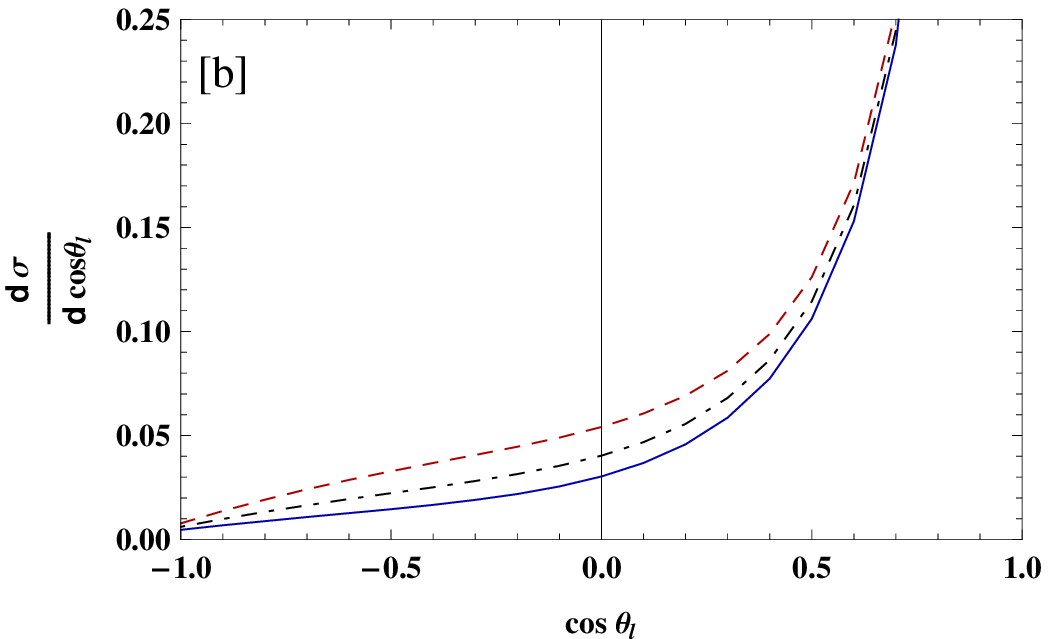}\\
\begin{center}
\includegraphics[width=6 cm,height=5.2 cm]{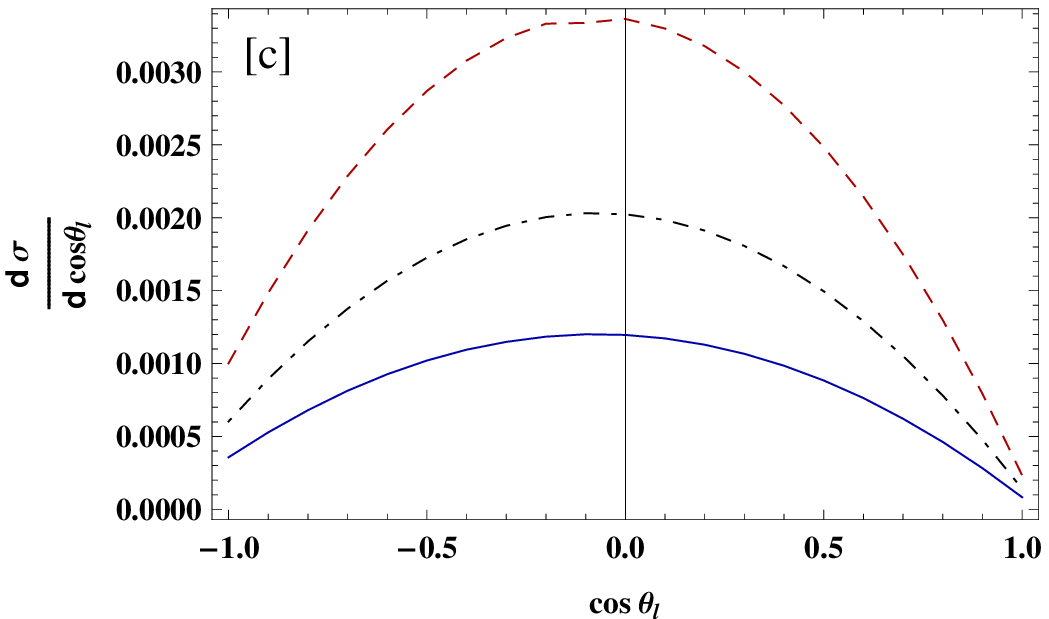}
\end{center}
\caption{Angular distribution of secondary leptons 
at $\sqrt{s}$=800 GeV within SM [blue - solid] and LHM with
 $f$=1 TeV, $c$=0.3 [red - dashed] and 
with $f$=1.5 TeV, $c$=0.3 [black - dot-dashed] for [a]  unpolarized beams,
[b] with $P_{e^{-}}$=-1 and $P_{e^{+}}$=0, and  
[c] with $P_{e^{-}}$= 1 and $P_{e^{+}}$=0}  
\label{fig:angdist}
\end{figure}

From the above energy-angle correlation, we obtain the $\cos\theta_{l}$ 
distribution by numerically integrating over $x.$ 
Fig.\ref{fig:angdist} shows the angular 
distribution for different polarization combinations for 
$\sqrt{s}=$ 800 GeV. 
It may be noted that the $\theta_l$ 
distribution closely follow the pattern of the angular distribution of the $W$,
which is expected to peak in the forward region for unpolarized beams and
with left-polarized
electron beams, but is symmetric in the case of right-polarized electron beams.
This is expected as the $W$ is produced with large kinetic energy, and the decay
leptons are expected to follow its momentum direction. 
The case of right-handed electron beam Fig.\ref{fig:angdist}$(c)$ 
is interesting. Recall that $Z_H$ couples only to the left-handed electrons.
Therefore, there is no contribution from the $Z_H$ exchange when we have
right-polarized electron beam. But, we still notice appreciable deviation
in the angular distribution compared to the SM case. This comes about through
the change in the SM couplings. Similar effect was seen in the 
case of total cross section also. 
In the case of $\sqrt{s}= 500$ GeV the effect is similar,
but somewhat less pronounced and therefore not displayed here explicitly.

A useful quantity 
to obtain in the case of unpolarized and left-polarized beams is the fraction
of leptons emitted in the backward direction, which may be defined as:
\[
f_{back}=\frac{\int_{-1}^0\left(d\sigma/d\cos\theta_l\right)~d\cos\theta_l}
{\int_{-1}^1\left(d\sigma/d\cos\theta_l\right)~d\cos\theta_l}\]
In Table~\ref{tab:f_back} we 
present these fractions for  
$\sqrt{s}=500$ GeV, and $800$ GeV.  The deviation is about 34 \% at 
$\sqrt{s}$ =500 GeV for $f$=1TeV and $c$=0.3 with unpolarised beams. 
Notice that the effect is not very sensitive to 
the choice of $c$.  But larger $f$ values tend to reduce the effect 
drastically. Using left-polarized electron beam ($P_{e^-}=-1$) does not 
affect the above results significantly. 
At $\sqrt{s}=800$ GeV, the effects are even more dramatic, as can
be judged from the Table.
In the case of right-polarized 
electron beam ($P_{e^-}=+1$), $f_{back}$ remains the same in both LHM 
as well as SM.

\begin{table}
\begin{center}
\begin{tabular}{|c|c|c|l|c|c|c|c|c|} \hline
&&& & &\multicolumn{2}{|c|}{}&\multicolumn{2}{|c|}{}\\[2mm] 
&&& & &\multicolumn{2}{|c|}{$\sqrt{s}$ =500GeV}&
\multicolumn{2}{|c|}{$\sqrt{s}$ =800GeV}\\[2mm] \cline{6-9}
&&&&&&&&\\
$P_{e^-}$ &$P_{e^+}$&Model& $f$ (TeV)&$c$
&$f_{back}$&$A_{FB}$ 
&$f_{back}$&$A_{FB}$\\[2mm] \hline\hline
  &  &SM & &      &0.035&-0.93&0.024&-0.95  \\[2mm]\cline{3-9}     
  &  &   & 1 &0.3   &0.047&-0.91&0.047&-0.91            \\
0 &0 &LHM&1.5&0.3   &0.040&-0.92&0.034&-0.93                \\
  &  &   &1.5&0.5   &0.039&-0.92&0.032&-0.94                \\ \hline
  &  &SM &   &      &0.032&-0.94&0.022&-0.96  \\[2mm]\cline{3-9}    
  &  &   & 1 &0.3   &0.044&-0.91&0.044&-0.91            \\
-1&0 &LHM&1.5&0.3   &0.037&-0.93&0.032&-0.94                \\
  &  &   &1.5&0.5   &0.037&-0.93&0.030&-0.94                \\ \hline
\end{tabular}
\end{center}
\caption{Fraction of leptons emitted 
in the backward direction, and the forward-backward asymmetry 
for both LHM and SM model for unpolarized and 
polarized beams with different choices of parameters.}
\label{tab:f_back}
\end{table}

Another useful observable related to the angular asymmetry is the 
forward-backward asymmetry defined as:
\begin{equation}
A_{FB}=\frac{\int_{-1}^0(d\sigma/d\cos\theta_l)~d\cos\theta_l-
             \int_{ 0}^1(d\sigma/d\cos\theta_l)~d\cos\theta_l}
            {\int_{-1}^1(d\sigma/d\cos\theta_l)~d\cos\theta_l}.
\end{equation}

In Table~\ref{tab:f_back} $A_{FB}$ is tabulated for different parameter
values at two different collider energies. Deviation of about 5\% is
observed in the asymmetry for $f=1$ TeV and $c=0.3$.

\subsection{Energy Spectrum of the Secondary Lepton}
The energy spectrum of the secondary leptons are sensitive to the $W^{\pm}$
helicities. The energy distribution in the centre of mass frame may be written
in terms of the polarization fractions of the $W$'s in 
the following form \cite{BESS}:
\begin{eqnarray}
\frac{1}{\sigma}\frac{d\sigma}{dx}  =   
\frac{2}{\beta^{3}}\left[\frac{3}{4}f^{0}(\beta^{2}-(1-2x)^{2})+
\right. & & \nonumber \\
 \left.
\frac{3}{8}f^{+}(\beta -1+2x)^{2}+\frac{3}{8}f^{-}(\beta +1-2x)^{2}\right].
& &
\end{eqnarray}
 
\begin{figure}[htb]
\centerline{\epsfysize=1.50 truein\epsfbox{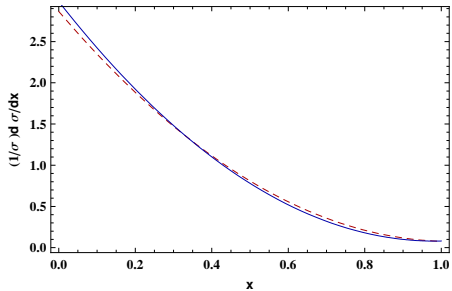}}
\caption{The energy spectrum of the charged decay lepton in 
$e^{+}e^{-} \rightarrow W^{+}W^{-}$,with 
$W^{-}\rightarrow l^{-}\bar{\nu}$ at $\sqrt{s}$=800 GeV.
Unpolarised beams are used for SM [blue - solid] and
LHM [red - dashed].
The parametres in this case are $f$=1 TeV and $c$=0.3}
\label{fig:Edist}
\end{figure}

In Fig.~\ref{fig:Edist} we present
the energy spectrum of the charged decay lepton for $\sqrt{s}=800$ GeV. We 
notice that a slightly larger fraction of hard leptons are produced in the
case of LHM compared to the case of the SM. The effect is much smaller in
the case of $\sqrt{s}=500$ GeV. While it is true that the effect is not 
dramatic, given the fact that the lepton energy spectrum could be obtained 
easily and with high efficiency, this observable might be useful in 
probing the LHM.

\subsection{Left-Right Asymmetry}
The new gauge boson $Z_{H}$ in the LHM has the peculiar property
of coupling only to left handed fermions as mentioned earlier.  
On the other hand, the SM $Z$ couples to both left- and right- handed fermions,
but the corrections to the $Ze^+e^-$ coupling is such that only the left-handed
electron coupling is affected. Thus, one would expect appreciable change in 
the asymmetry between the left- and right-polarized cross sectionos. 

\begin{figure}[htb]
\includegraphics[width=6 cm,height=5 cm]{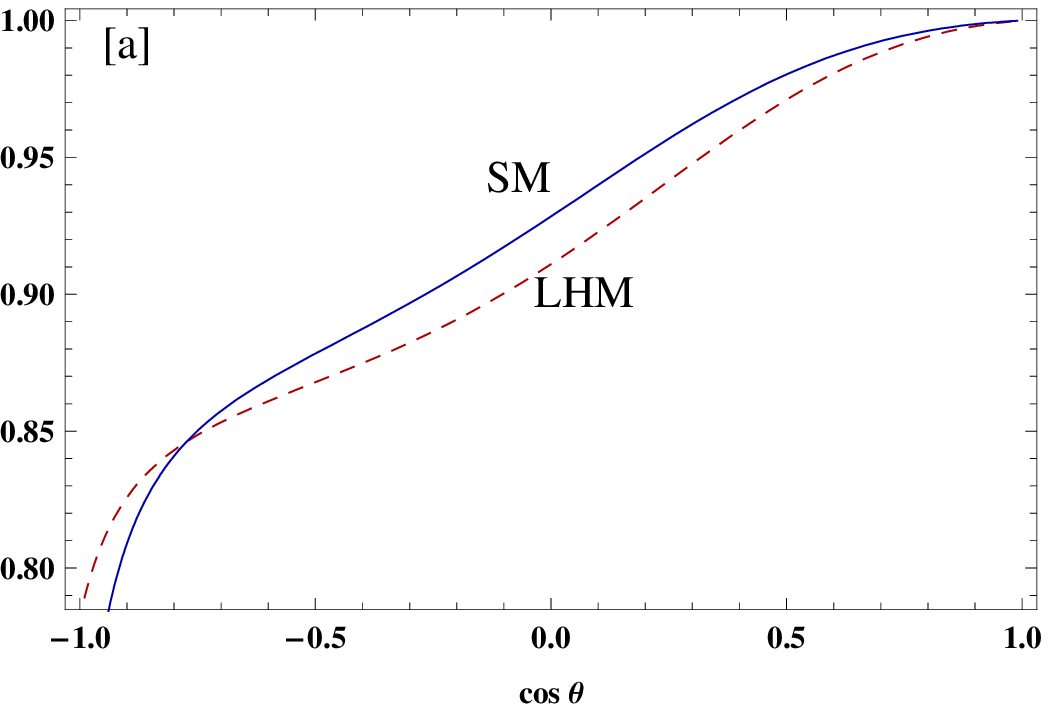}
\hspace{0.2cm}
\includegraphics[width=6 cm,height=5 cm]{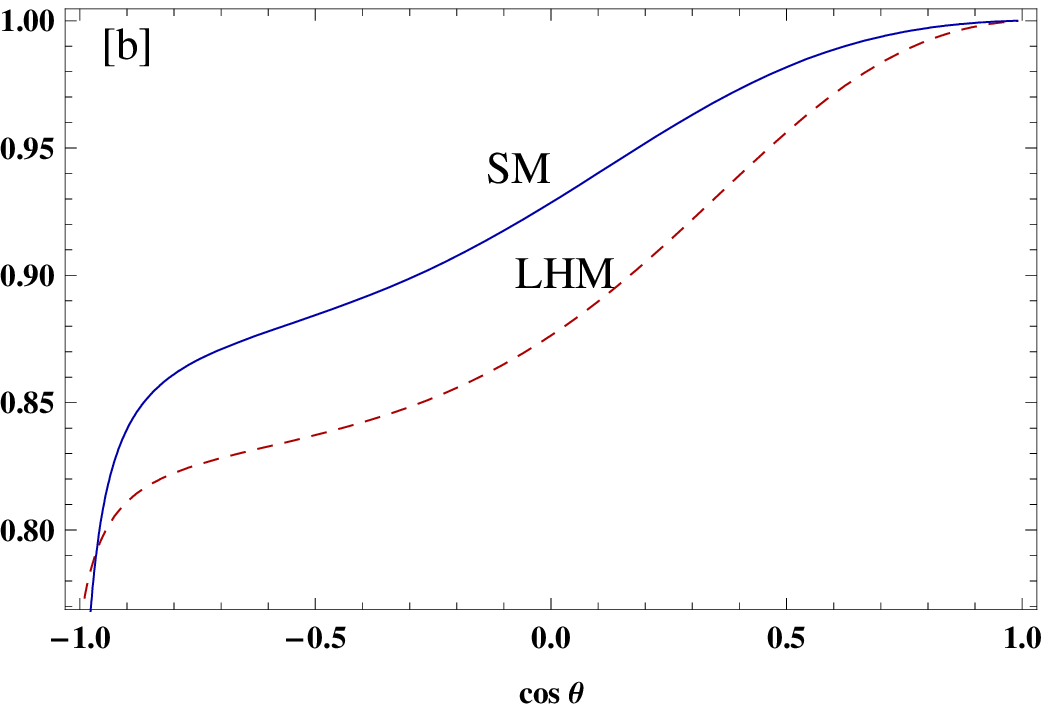}
\caption{The left-right asymmetry for SM [blue - solid]
 and LHM [red - dashed] at [a]$\sqrt{s}$=500GeV [b]$\sqrt{s}$=800GeV. 
The parametres considered are $f=1$ TeV and 
$c=0.3$.}
\label{fig:lr-asym}
\end{figure}

We define the left-right asymmetry in the differential cross section as:
\begin{equation}
A^{diff}_{LR}=\frac{(d\sigma(e^-_Le^+_R) / d\cos\theta-d\sigma(e^-_Re^+_L) / 
d\cos\theta)}
{(d\sigma(e^-_Le^+_R) / d\cos\theta+d\sigma(e^-_Re^+_L) / d\cos\theta)},
\end{equation}
where $\theta$ is the $W$ scattering angle.
Fig.~\ref{fig:lr-asym} shows the LR asymmetry for two energies.
Even at low energies the deviation becomes apparent. 

\begin{table}
\begin{center}
\begin{tabular}{|l|l|c|c|c|c|c|} \hline
&&&\multicolumn{2}{|c|}{$A_{LR}$} \\ \cline{4-5}
Model&$f$ (TeV)&$c$&$\sqrt{s}$ =500GeV&$\sqrt{s}$ =800GeV \\ 
\hline
\multicolumn{5}{l}{}\\
\multicolumn{5}{l}{$\theta_0=0$}\\
\hline
\multicolumn{3}{|l|}{SM}      &0.992 & 0.995 \\\hline     
& 1 &0.3  &0.988 & 0.986            \\
LHM&1.5&0.3  & 0.990 & 0.992                \\
&1.5&0.5  & 0.991 & 0.993                 \\ \hline
\multicolumn{5}{l}{}\\
\multicolumn{5}{l}{$\theta_0=15^o$}\\
\hline
\multicolumn{3}{|l|}{SM}      &0.985 & 0.987 \\\hline     
& 1 &0.3  &0.978 & 0.967            \\
LHM&1.5&0.3  & 0.982 & 0.979                \\
&1.5&0.5  & 0.983 & 0.981                 \\ \hline
\multicolumn{5}{l}{}\\
\multicolumn{5}{l}{$\theta_0=30^o$}\\
\hline
\multicolumn{3}{|l|}{SM}      &0.974 & 0.976 \\\hline     
& 1 &0.3  &0.962 & 0.945            \\
LHM&1.5&0.3  & 0.969 & 0.964                \\
&1.5&0.5  & 0.970 & 0.967                 \\ \hline
\multicolumn{5}{l}{}\\
\multicolumn{5}{l}{$\theta_0=45^o$}\\
\hline
\multicolumn{3}{|l|}{SM}      &0.960 & 0.962 \\\hline     
& 1 &0.3  &0.945 & 0.920            \\
LHM&1.5&0.3  & 0.953 & 0.945                \\
&1.5&0.5  & 0.955 & 0.949                 \\ \hline
\multicolumn{5}{l}{}\\
\multicolumn{5}{l}{$\theta_0=60^o$}\\
\hline
\multicolumn{3}{|l|}{SM}      &0.944 & 0.946 \\\hline     
& 1 &0.3  &0.927 & 0.897            \\
LHM&1.5&0.3  & 0.937 & 0.926                \\
&1.5&0.5  & 0.939 & 0.931                 \\ \hline
\end{tabular}
\end{center}
\caption{ $A_{LR}$ for various opening angles $\theta_0$
for SM and LHM 
with different choice of parameters.}
\label{tab:asym}
\end{table}
We may go one step further by considering an integral version of
this asymmetry as better efficiency may be obtained this way, 
by integrating each of the differential cross sections
from an opening angle $\theta_0$ up to an angle $\pi-\theta_0$,
for various realistic values of $\theta_0$ to which the data
can be integrated without difficulty.  
We define the integrated left-right asymmetry as:
\begin{equation}
A_{LR}=\frac{\sigma_{\theta_0}(e^-_Le^+_R\rightarrow W^+W^-)-
\sigma_{\theta_0}(e^-_Re^+_L\rightarrow W^+W^-)}
{\sigma_{\theta_0}(e^-_Le^+_R\rightarrow W^+W^-)+
\sigma_{\theta_0}(e^-_Re^+_L\rightarrow W^+W^-)}
\end{equation}
where $\sigma_{\theta_0}$ stands for $\int_{\theta_0}^{\pi-\theta_0}
\left( d\theta~(d\sigma/d\theta\right)$.
This asymmetry, for different parameter values at $\sqrt{s}=500$ GeV and
at $\sqrt{s}=800$ GeV is tabulated in Table ~\ref{tab:asym}. 
We see that the asymmetry is not affected in any significant
way at $\sqrt{s}=500$ GeV or at $\sqrt{s}=800$ GeV for the parameter
combinations considered, when no cut on angle is applied. 
Interesting patterns may be observed from this table.  An interplay
between the differential asymmetry plotted above, and the fact 
that the bulk of the contribution to the cross section comes
from the forward region where the asymmetry itself is not appreciable
leads to more sigificant results when the cut-off angle is larger.
In other words, as the cut-off angle increases, the region of
deviation between the LHM and the SM is weighted more efficiently.
The case of $\theta_0=15^o$ is worthy of note, as there is a cross-over
in the asymmetry for $\sqrt{s}=800$ GeV and the effect is completely
is wiped out.

\section{Conclusions}
Understanding the phenomenon of electroweak symmetry breaking is central to 
the study of elementary particle physics. Among the viable alternatives to the 
Standard Higgs Mechanism is the Little Higgs Scenarios, which provides a 
natural way to dynamically generate electroweak symmetry breaking. In 
this report we have considered
the LHM which is one simple version of this scenario. 
Such a model predicts the existence of additional
gauge bosons with masses in the TeV region. 
We have considered the process $e^{+}e^{-}\rightarrow W^{+}W^{-}$ 
to probe this model. Presence of a heavy neutral gauge boson in addition
to the SM gauge bosons, and the change of couplings of the SM particles 
affect this process. 

We have studied the total cross section and the polarization fraction of the 
$W$'s produced for typical parameter values. We conclude that for suitable
choice of beam polarizations, there can be more than 50\% deviation in 
the cross section for $\sqrt{s}$ around and above 500 GeV. The polarization 
fractions are found to be more sensitive to the new effects, which can 
be two or three times the SM value at $\sqrt{s}=800$ GeV. The fact that
the $W$ polarization fractions can be very precisely measured at ILC shows
that study of these observables can effectively probe the 
LHM. 

Study of secondary lepton distributions can be carried out efficiently
at the ILC. Our study of the energy and angular 
distributions in the laboratory frame shows that significant deviation from the 
SM expectation is possible for parameter sets of LHM allowed by the present
experimental constraints. We see that the angular distribution is better
suited in the present case with fraction of secondary leptons emitted in 
the backward direction deviates from the SM value by significant amounts.
We have introduced a LR asymmetry for the differential as well as
the integrated cases, where the integration is performed over an
opening angle given by $\theta_0$.  It is shown that a judicious
choice of $\theta_0$ can provide a window for observing striking deviations
from the SM and provide a discriminating tool for the LHM.

While a more complete study including,
for instance, detector efficiencies, and larger
parameter space scan is pending, our 
study with representative parameter points illustrates that $W$ pair
production in $e^+e^-$ collisions at ILC energies can probe the LHM
very effectively. 

\bigskip

\noindent
\textbf{Acknowledgements:} 
BA thanks the Department of Science and Technology,
Government of India for support during the course of these investigations.

\bigskip

\noindent
\textbf{\large Appendix:  $A(s,x,\theta_l)$}

\bigskip

\noindent
The expression for energy-angle correlation for the unpolarised beams 
is given in Ref.~\cite{Kov, BESS}, for the SM and for the BESS models
respectively, which may be easily adapted for the LHM.  The
generalization to include arbitary beam polarization
for each of these models can also be done in a straightforward
manner.  Here we give it for the LHM.    
The expressions result from straightforward Dirac trace techniques to
include longitudinal beam polarization, by evalauting the electronic part of
the Feynman diagrams given in Fig. 1.  In particular, for $C_s$ and $C_{s}'$
the requisite traces which produce the beam polarization dependence
associated with them are analogous to those
that have been explicitly discussed in
ref.~\cite{BASDR}.
The beam polarization dependence in $C_{int}$ and $C_t$ which we have
presented explicitly here, reflect the maximal parity violating nature of
the couplings of the Ws to the electrons and positrons in the t-channel
diagram.

For arbitrary beam 
we have $A(s,x,\theta_{l})=C_{s}A_{s}+C^{'}_{s}A^{'}_{s}+C_{int}A_{int}+
C_{t}A_{t},$ with
\begin{eqnarray}
A_{s}&=&-\frac{3}{2}-\tau-\frac{\tau}{x}+\frac{\tau^{2}}{x^{2}}+\frac{x}{\tau}(1-x)
\left(1+\frac{1}{4\tau}\right) \nonumber\\
&&+\left(-\frac{5}{2}-\tau+3\frac{\tau}{x}-3\frac{\tau^{2}}{x^{2}}+\frac{1}{2\tau}
+\frac{x}{\tau}(1-x)\left(1-\frac{1}{4\tau}\right)\right)\cos^{2}\theta_{l},\nonumber\\[2mm]
A^{'}_{s}&=&2\left(1+\frac{1}{4\tau}-2x-2\frac{\tau}{x}\right)\cos\theta_{l},\nonumber\\[2mm]
A_{int}&=&-2\tau+\frac{x}{\tau}-2+\frac{x}{2\tau}(1-x)\left(1+\frac{1}{2\tau}\right)
+\left(1+\frac{1}{2\tau}-\frac{2\tau}{x}-2x\right)\cos\theta_{l}\nonumber\\
&&-\left(1-\frac{1}{2\tau}\right)\left(1-\frac{x(1-x)}{2\tau}\right)\cos^{2}\theta_{l}\nonumber\\
&&-Rx^{2}\left(2+(\cos\theta_{l}-\beta\cos\theta)
\times\left(2-\left(1+\frac{1}{\tau}\right)\beta\cos\theta+\cos\theta_{l}\right)\right),\nonumber \\[2mm] 
A_{t}&=&\left(-2+\frac{2x}{\tau}+\frac{x(1-x)}{4\tau^{2}}\right)+\frac{\cos\theta_{l}}{2\tau}
+\left(1-(1-x)\frac{x}{2\tau}\right)\frac{\cos^{2}\theta_{l}}{2\tau}\nonumber\\
&&-\frac{2}{\tau}x^{2}R(\beta\cos\theta-\cos\theta_{l})\beta\cos\theta
+2x^{2}aR^{3}(\beta\cos\theta-\cos\theta_{l})^{2}. \nonumber
\end{eqnarray}
Here
\begin{eqnarray}
&&R= \left[4\tau^{2}+(\beta\cos\theta-\cos\theta_{l})
(\beta\cos\theta-\beta^{2}\cos\theta_{l})\right]^{-\frac{1}{2}}\nonumber\\[2mm] 
&&a=2\tau-1+\beta\cos\theta\cos\theta_{l}, \nonumber
\end{eqnarray}
where $\cos\theta={1}/{\beta}\left(1-{2\tau}/{x}\right)$ is the 
scattering angle of $W^-$ and $\beta=(1-4m^{2}_{W}/s)^{\frac{1}{2}}$ is the
velocity of $W^-$, both in the centre of mass frame. 
The coefficients, $C$'s are 
given below for arbitrary beam polarizations with $P_{e^-}$ and $P_{e^+}$ 
denoting the degrees of electron and positron beams, respectively. 
In addition
to the standard channels (as given in Ref.~\cite{Kov, BESS}), these 
coefficients include the contribution due the $Z_H$ exchange.

\begin{eqnarray}
C_{s}&=&\left(1-P_{e^{-}}P_{e^{+}}\right)~\left(g^{2}_{\gamma WW}\left[(c^{v}_{\gamma})^{2}+(c^{a}_{\gamma})^{2}-P~(2 c^{a}_{\gamma} c^{v}_{\gamma})\right]\right.+\nonumber \\ 
 && s^{2}_{Z}~g^{2}_{ZWW}~\left[(c^{v}_{Z})^{2}+(c^{a}_{Z})^{2}-P~(2 c^{a}_{Z}c^{v}_{Z})\right]+\nonumber \\ 
      && s^{2}_{Z_{H}}~g^{2}_{Z_{H}WW}~\left[(c^{v}_{Z_{H}})^{2}+(c^{a}_{Z_{H}})^{2}-P~(2 c^{a}_{Z_{H}}c^{v}_{Z_{H}})\right]+ \nonumber \\
&& 2~ s_{Z}~g_{\gamma WW}~g_{ZWW}~\left[c^{v}_{\gamma}c^{v}_{Z}+c^{a}_{\gamma}c^{a}_{Z}-P~(c^{a}_{\gamma}c^{v}_{Z}+c^{a}_{Z}c^{v}_{\gamma})\right]+\nonumber  \\        
&& 2~s_{Z_{H}}~g_{\gamma WW}~g_{Z_{H}WW}~\left[c^{v}_{\gamma}c^{v}_{Z_{H}}+c^{a}_{\gamma}c^{a}_{Z_{H}}-P~(c^{a}_{\gamma}c^{v}_{Z_{H}}+c^{a}_{Z_{H}}c^{v}_{\gamma})\right]+\nonumber \\ 
&& 2~s_{Z}~s_{Z_{H}}~g_{ZWW}~g_{Z_{H}WW}~\left.\left[c^{v}_{Z}c^{v}_{Z_{H}}+c^{a}_{Z}c^{a}_{Z_{H}}-P~(c^{a}_{Z}c^{v}_{Z_{H}}+c^{a}_{Z_{H}}c^{v}_{Z})\right]\right)\nonumber\\[2mm]
C^{'}_{s}&=&(1-P_{e^{-}}P_{e^{+}})~\left(g^{2}_{\gamma WW}~\left[2c^{a}_{\gamma}c^{v}_{\gamma}-P~((c^{v}_{\gamma})^{2}+(c^{a}_{\gamma})^{2})\right]\right.+\nonumber \\ 
&& s^{2}_{Z}~g^{2}_{ZWW}~\left[2c^{a}_{Z}c^{v}_{Z}-P((c^{v}_{Z})^{2}+(c^{a}_{Z})^{2})\right]+ \nonumber \\
&& s^{2}_{Z_{H}}~g^{2}_{Z_{H}WW}~\left[2c^{a}_{Z_{H}}c^{v}_{Z_{H}}-P~((c^{v}_{Z_{H}})^{2}+(c^{a}_{Z_{H}})^{2})\right]+\nonumber \\ 
&& 2~s_{Z}~g_{\gamma WW}~g_{ZWW}~\left[(c^{a}_{Z}c^{v}_{\gamma}+c^{a}_{\gamma}c^{v}_{Z})-P~(c^{v}_{\gamma}c^{v}_{Z}+c^{a}_{\gamma}c^{a}_{Z})\right]+\nonumber \\ 
&& 2~s_{Z_{H}}~g_{\gamma WW}~g_{Z_{H}WW}~\left[(c^{a}_{Z_{H}}c^{v}_{\gamma}+c^{a}_{\gamma}c^{v}_{Z_{H}})-P~(c^{v}_{\gamma}c^{v}_{Z_{H}}+c^{a}_{\gamma}c^{a}_{Z_{H}})\right]+ \nonumber \\ 
&& 2~s_{Z}~s_{Z_{H}}~g_{Z_{H}WW}~g_{ZWW}~\left.\left[(c^{a}_{Z_{H}}c^{v}_{Z}+c^{a}_{Z}c^{v}_{Z_{H}})-P~(c^{v}_{Z}c^{v}_{Z_{H}}+c^{a}_{Z_{H}}c^{a}_{Z})\right]\right)\nonumber\\[2mm]
C_{int}&=&g^{2}_{e\nu W}~\left(g_{\gamma WW}~(c^{v}_{\gamma}+c^{a}_{\gamma})+s_{Z}~g_{ZWW}~(c^{v}_{Z}+c^{a}_{Z})+s_{Z_{H}}~g_{Z_{H}WW}~(c^{v}_{Z_{H}}+c^{a}_{Z_{H}})\right) \nonumber \\ 
&&\times(1-P_{e^{-}})(1+P_{e^{+}})\nonumber\\[2mm]
C_{t}&=&\frac{g^{4}_{e\nu W}}{2}\times(1-P_{e^{-}})(1+P_{e^{+}})\nonumber
\end{eqnarray}
Here, the effective polarization,
$P=(P_{e^{-}}-P_{e^{+}})/(1-P_{e^{-}}P_{e^{+}})$, and the propagator factors 
are defined as: $s_V={s}/{(s-m_V^2)},$ where $m_V$ is the mass of the 
corresponding gauge boson, $V=Z,~Z_H$. We have assumed that the centre of mass
energy is sufficiently
far away from the threshold regions of the gauge bosons involved.

\end{document}